\documentclass[11pt]{article}

\usepackage[utf8]{inputenc}
\usepackage{authblk}

\usepackage{graphicx}
\usepackage[margin=1.0in]{geometry}
\usepackage{braket}
\usepackage{amsmath}
\usepackage{amsfonts}
\usepackage{float}
\usepackage{url}
\usepackage{comment} 
\usepackage{caption}
\usepackage{subcaption}
\usepackage{booktabs}
\usepackage[colorlinks]{hyperref}
\usepackage{lineno}

\usepackage{colortbl}

\definecolor{Lightgray}{RGB}{235,235,235}

\title{Hierarchical quantum classifiers}

\author[1]{Edward Grant\thanks{\texttt{\href{mailto:edward.grant.16@ucl.ac.uk}{edward.grant.16@ucl.ac.uk}}}}
\author[1,2]{Marcello Benedetti}
\author[3,4]{Shuxiang Cao}
\author[5,6]{Andrew Hallam}
\author[1]{\par Joshua Lockhart}
\author[7]{Vid Stojevic}
\author[5]{Andrew G. Green}
\author[1]{Simone Severini}

\affil[1]{Department of Computer Science, University College London, WC1E 6BT London, UK}
\affil[2]{Cambridge Quantum Computing Limited, CB2 1UB Cambridge, UK}
\affil[3]{College of Computer Science and Technology, Zhejiang University, 310027 Hangzhou, China}
\affil[4]{Clarendon Laboratory, Department of Physics, University of Oxford, OX1 3PU Oxford, UK}
\affil[5]{London Centre for Nanotechnology, 17-19 Gordon St, London, WC1H 0AH London, UK}
\affil[6]{Department of Physics and Astronomy, University College London, WC1E 6BT London, UK}
\affil[7]{GTN Limited, EC1V 2PY London, UK }

\date{\normalsize\today}

\begin{document}

\maketitle

\begin{abstract}
Quantum circuits with hierarchical structure have been used to perform binary classification of classical data encoded in a quantum state. We demonstrate that more expressive circuits in the same family achieve better accuracy and can be used to classify highly entangled quantum states, for which there is no known efficient classical method. We compare performance for several different parameterizations on two classical machine learning datasets, Iris and MNIST, and on a synthetic dataset of quantum states. Finally, we demonstrate that performance is robust to noise and deploy an Iris dataset classifier on the ibmqx4 quantum computer.
\end{abstract}

\section{Introduction} 

Neural networks offer state-of-the-art performance in a wide number of machine learning tasks including computer vision, natural language processing, generative modelling and reinforcement learning~\cite{lecun2015deep}. The hierarchical structure of deep neural networks can allow them to match the expressiveness of shallower models with exponentially fewer parameters~\cite{lin2017does, mhaskar2016learning, telgarsky2016benefits}. In recent years, there has been much interest in translating the success of neural networks to the quantum computing context~\cite{schuld2014quest}. Despite this, there are many open questions regarding the advantages that quantum computation can bring to machine learning~\cite{Ciliberto20170551, aaronson2015read}. Do quantum algorithms offer a clear speed-up over classical approaches for inference and training? Are quantum machine learning algorithms robust to noise? What are the best quantum circuit layouts to carry out machine learning tasks? An exciting way to explore these questions is through experimentation on available quantum hardware, and simulation on classical hardware. 

Tensor networks are a method for representing an intractable high rank tensor as a decomposition of tractable lower rank tensors connected by contraction. They are widely used in many-body physics for the simulation of strongly correlated quantum systems, and can be used to represent both quantum states and quantum circuits~\cite{2014AnPhy.349..117O,shi2006classical,vidal2008class,verstraete2008matrix}. Tensor networks with hierarchical structure exhibit many similarities with neural networks and in some cases have been shown to be equivalent~\cite{cohen2016convolutional,levine2018bridging}. Given that tensor networks can be used to represent both neural networks and quantum circuits, they are a natural choice for exploring the intersection of both fields. In this work, we consider the supervised machine learning tasks of classifying classical and quantum data on a quantum computer using hierarchical quantum circuits. 

To perform classification on a quantum computer the input data must be encoded in a quantum state. Two ways in which this can be achieved are by encoding the data in the amplitudes of individual qubits in a fully separable state (qubit encoding), or in the amplitudes of an entangled state (amplitude encoding). We test classifiers using both encoding methods. For classical data we perform qubit encoding using single qubit rotations. For quantum data we assume that the data arrives from another quantum device and is already an entangled amplitude encoded state.  

Once the data has been encoded, the classifier consists of a series of unitary operations applied to the initial quantum state. Then, a measurement is carried out on a target qubit. In practice, multiple runs are required to approximate the expectation of the measurement outcome, and the most frequent outcome is taken as the predicted class. More runs increase the classifier confidence.

In addition to the pipeline just described, we need to specify the layout of the hierarchical circuit, and the algorithm for learning its parameters. The circuits we use here are tree-like and can be parameterized with a simple gate-set that is compatible with currently available quantum computers. The first of these circuits is known as a Tree Tensor Network (TTN)~\cite{shi2006classical}. We then consider a more complex circuit layout known as the Multi-Scale Entanglement Renormalization Ansatz (MERA)~\cite{vidal2008class}. MERAs are similar to TTNs, but make use of additional unitary transformations to effectively capture a broader range of quantum correlations. Both one-dimensional ($1$D) and two-dimensional ($2$D) versions of TTN and MERA circuits have been proposed in the literature~\cite{cincio2008multiscale,evenbly2010entanglement}.

In the $1$D case, TTN and MERA circuits can be evaluated efficiently using classical techniques when the input data is qubit encoded. Evaluating such circuits on amplitude encoded data is likely to be classically intractable. In $2$D, the TTN circuit is efficiently simulatable when using qubit encoding, whereas the $2$D MERA circuit is not. Because we cannot simulate large $2$D MERA circuits, we restrict our experiments to the $1$D case. In all experiments we find that $1$D MERA outperforms $1$D TTN, suggesting that $2$D MERA could in principle outperform $2$D TTN. Such a hypothesis should be tested with future experiments as suitably large near-term quantum computers become available. Classifiers that possess a $1$D structure could be used for sequential data such as time-series, while classifiers that possess a $2$D structure would be the natural choice for $2$D data such as a natural images.

Optimizing the circuits can be accomplished by stochastic gradient descent. In the case of efficiently simulatable networks, it makes sense to use the analytic gradient. For circuits that cannot be efficiently simulated it is possible to use a quantum computer and estimate gradients numerically. Moreover, a hybrid approach that involves a classical pre-training step to initialize some of the gates has been previously proposed~\cite{2018arXiv180311537H}. We empirically validate this approach by initializing a $1$D MERA with a pre-trained $1$D TTN. Such pre-training reduces the average number of training steps needed until convergence on a model with comparable accuracy, a benefit for implementations on near-term quantum computers.

We demonstrate our techniques using TTNs and MERAs and compare performance for a number of parameterizations. The first of these uses only single-qubit rotations and fixed CNOT gates. The second uses more general two-qubit gates. The third uses three-qubit gates, where the additional ancilla qubits allow for non-linear operations. Both real and complex parameterizations are compared. 

We test the ability of each classifier to predict binary labels on two canonical machine learning datasets, Iris~\cite{fisher1936use} and MNIST handwritten digits~\cite{lecun1998mnist}, and on synthetic quantum datasets. We also use the IBM Quantum Experience~\cite{ibmqx} to test robustness to depolarizing noise, and to deploy the model on the ibmqx4 quantum computer. 

The structure of the article is the following. Section.~\ref{sec:main_section} contains a description of the hierarchical quantum classifiers and the results of experiments on classical data and quantum states. Section~\ref{sec:discussion} contains a discussion of the results and a comparison to existing methods. Section~\ref{sec:conclusions} presents our conclusions and directions for future work.

\section{Results}\label{sec:main_section}

\subsection{Data encoding}

Classification consists of assigning a category to an observation. In machine learning, an inference model is trained to minimize the classification error on a finite set of data, also known as the training set. The actual performance of the classifier, the generalization error, is then estimated on a set of data points not used for training, also known as the test set. The functional form of the inference model is often critical to the success of the classifier. State-of-the-art models for high-dimensional datasets with complex structure are typically hierarchical or compositional~\cite{lecun2015deep}. These ideas can be translated to the paradigm of quantum computation using the framework of tensor networks. Before describing the tensor network architectures used in this work, namely TTN and MERA, it is important to first clarify what datasets are considered in this paper to gauge the performance of these networks, and how they are prepared. 

Let us first consider the case of classical data. A classical dataset for binary classification is a set $\mathcal{D} = \{(\boldsymbol{x}^{d}, y^d)\}_{d=1}^D$, where $\boldsymbol{x}^{d} \in \mathbb{R}^N$ are $N$-dimensional input vectors, and $y^{d} \in \{0, 1\}$ are the corresponding class labels. Classifying classical data on a quantum computer requires that the input vectors be encoded in a quantum state. There are a variety of ways to accomplish this and different algorithms require different encoding methods. The most efficient approach in terms of space is to encode classical data in the amplitudes of a superposition, that is, using $N$ qubits to encode a $2^{N}$ dimensional data vector. However, in the general case and depending on the quantum classifier used, the computational cost of preparing data as a superposition can negate the speedup obtained during classification~\cite{aaronson2015read}. A simpler method is to encode each element of a classical data vector in the amplitude of a single-qubit. This type of encoding requires $N$ qubits to encode an $N$ dimensional data-vector and, therefore, is less efficient in terms of space. However, the state preparation is clearly efficient in terms of time as it only requires single-qubit rotations. We opt for this type of encoding for classical data. In particular, we first re-scale the data vectors element-wise to lie in $[0, \frac{\pi}{2}]$. Then, we encode each vector element in a qubit using the following scheme~\cite{stoudenmire2016supervised}:
\begin{equation}
\psi_{n}^d=\cos(x_{n}^d)\ket{0}+ \sin(x_{n}^d)\ket{1} .
\label{e:encoding}
\end{equation}
The final data vector writes as $\psi^d = \otimes_{n=1}^N \psi_n^d$, and is ready to be used in a quantum algorithm.

Let us now consider the case of quantum data. A quantum dataset for binary classification is a set $\mathcal{D} = \{\psi^{d}, y^d)\}_{d=1}^D$, where $\psi^{d} \in \mathbb{C}^{2^N}$ are $2^N$-dimensional input vectors of unit length, and $y^{d} \in \{0, 1\}$ are the corresponding classes. In contrast to classical data, quantum data such as the output of a quantum circuit or a quantum sensor, may already be in superposition. That is, the quantum states are used as-is, and there is no relevant cost for the preparation.

\subsection{Circuit architecture}

\begin{figure}
    \centering
    \begin{subfigure}[b]{0.495\textwidth}
       \includegraphics[width=\textwidth]{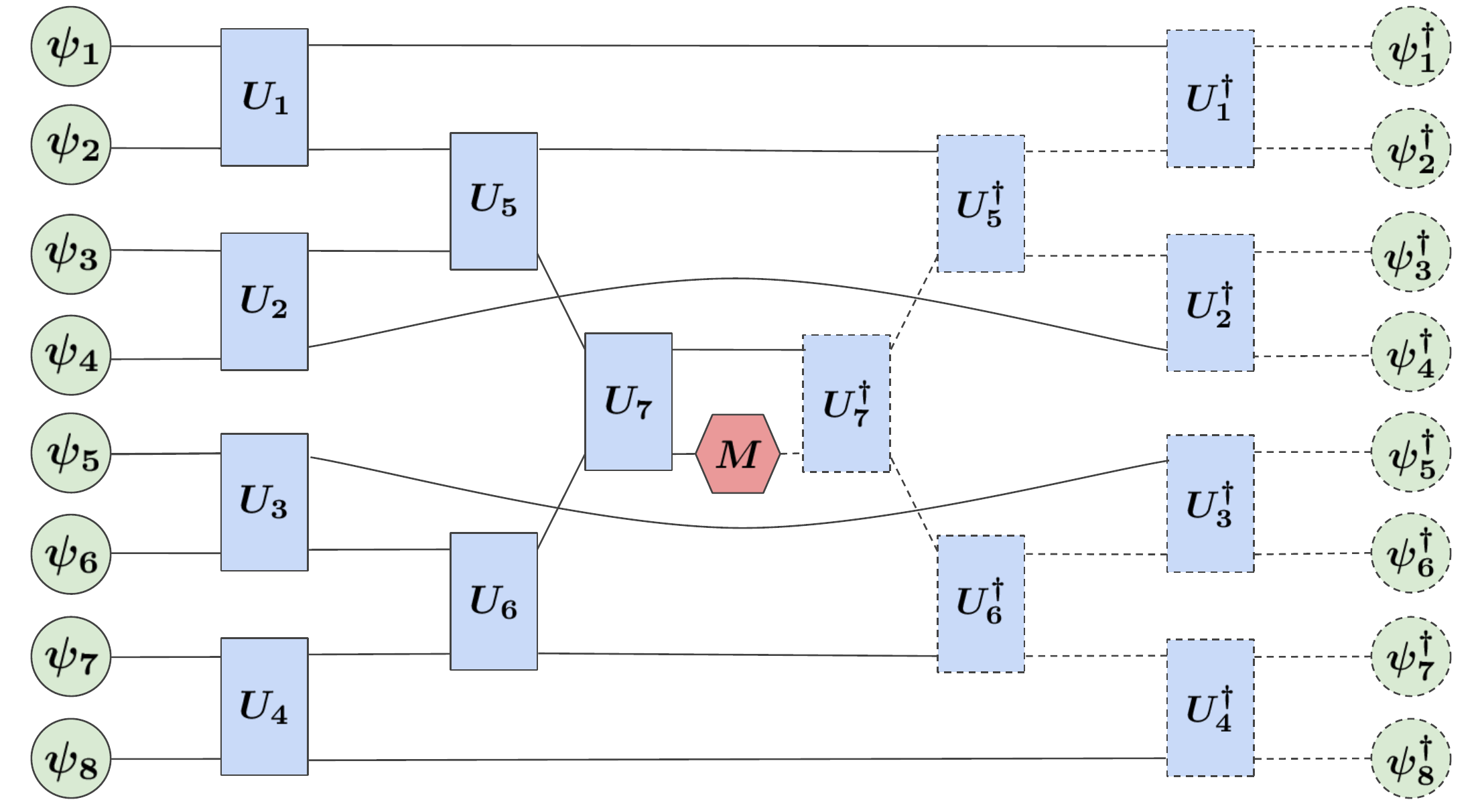}
       \caption{TTN classifier}
       \label{fig:TTN}
    \end{subfigure}
    \begin{subfigure}[b]{0.495\textwidth}
       \includegraphics[width=\textwidth]{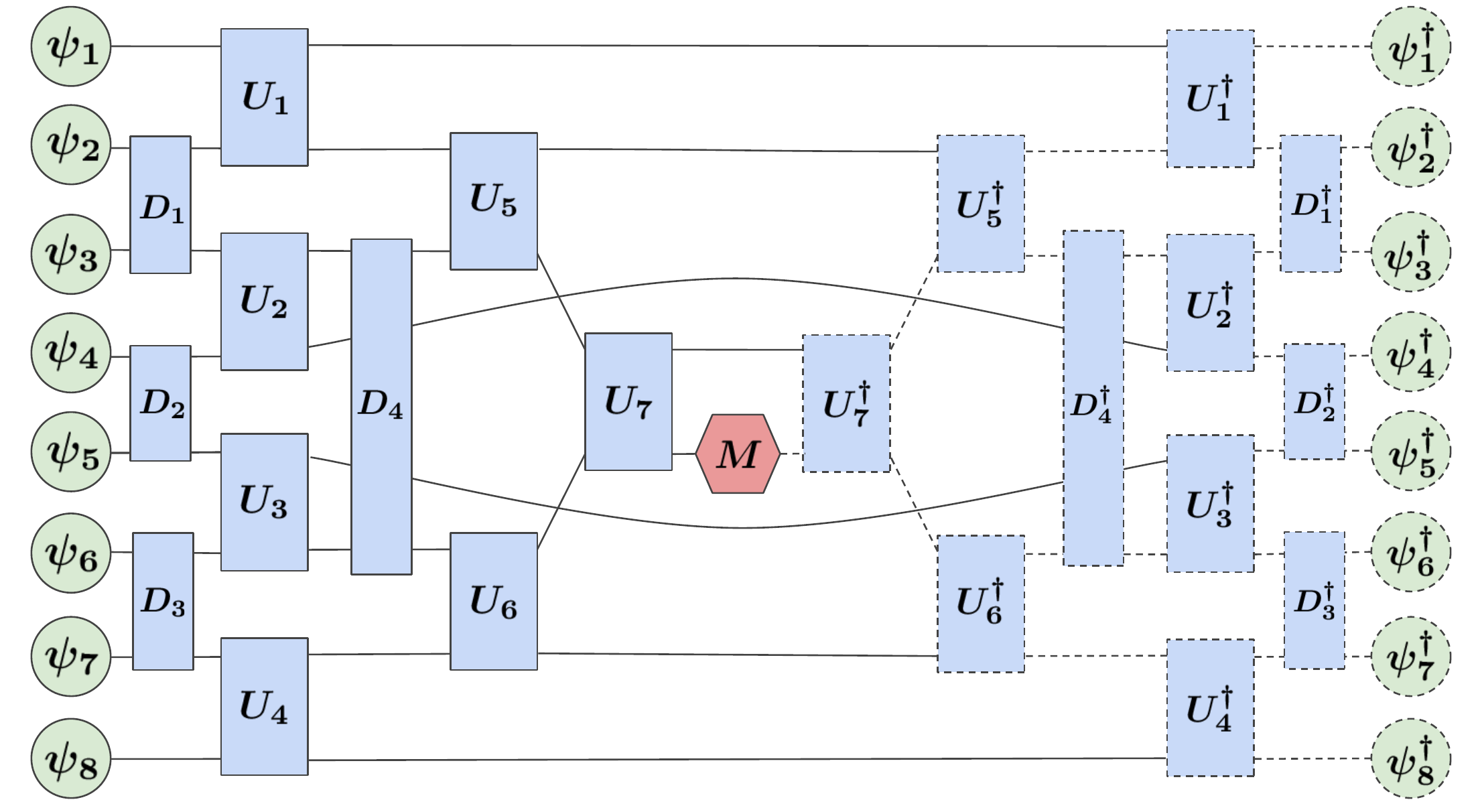}
       \caption{MERA classifier}
       \label{fig:MERA}
    \end{subfigure}
    \caption{\textit{TTN and MERA classifiers for 8 qubits.} The quantum circuit is illustrated by the regions outlined in solid lines comprising inputs $\psi$, unitary blocks $\{U_i\}_{i=1}^7$ and $\{D_i\}_{i=1}^4$, and a measurement operator $M$. The dashed lines represent its conjugate transpose. The solid and dashed regions together describe a tensor network operating on input $\psi_{1-8}$ and evaluating to the expectation value of observable $M$.}
    \label{fig:networks}
\end{figure}

We now discuss the quantum circuit architectures for classification. The first circuit architecture is inspired by tree tensor networks, specifically binary trees. The TTN circuit begins by applying a set of two-qubit nearest neighbour unitaries to the input. We then discard one of the qubits output from each unitary, halving the number of qubits in the next layer of the circuit. In the following layer we again apply two-qubit unitaries to the remaining qubits before discarding half of them. This process is repeated until only one qubit remains. The network in full consists of measuring a single-qubit expectation value on this remaining qubit
\begin{equation}
M_{\boldsymbol{\theta}}(\psi^d)=\langle \psi^d|\hat{U}_{QC}^\dagger(\{U_i(\theta_i)\})\hat{M} \hat{U}_{QC}(\{U_i(\theta_i)\})| \psi^d\rangle ,
\end{equation}
where $\hat{U}_{QC}(\{U_i\})$ is the quantum circuit made up of unitaries $U_i(\theta_i)$, $\boldsymbol{\theta}=\{\theta_i\}$ is the set of parameters which define the unitaries, and $\hat{M}$ is the single-qubit operator whose expectation we are calculating. A circuit diagram of an 8-qubit TTN is shown in Fig.~\ref{fig:networks}~(a). The solid lines encompass the circuit, while the dashed lines represent its conjugate transpose.

The MERA network is closely related to the TTN. All of the unitaries that make up a tree network are maintained with an additional layer of two qubit unitaries added before each layer of the TTN. These additional unitaries, $\{D_i\}$, each operate on one qubit of neighbouring unitaries in the upcoming TTN layer. In a conventional MERA network, the addition of these unitaries allows quantum correlations on a particular length scale to be captured at the same layer of the network ~\cite{vidal2008class}. A circuit diagram of an 8-qubit MERA is shown in Fig.~\ref{fig:networks}~(b).


\subsection{Unitary parameterization}

We have explored a number of different ways to parameterize the unitaries used in these circuits. Some of the input data used is purely real, we therefore tested the effect of restricting the unitaries to be real too. That is, we chosen unitaries such that $U_i \in SO(\cdot) \subset SU(\cdot)$. We also consider general, complex valued unitaries $U_i \in  SU(\cdot)$. As has been observed in the context of the time-dependant variational principle applied to tensor networks, the use of complex weights often prevents optimization from getting stuck in local minima~\cite{haegeman2011time,trabelsi2017deep}.

\begin{figure}
    \centering
    \begin{subfigure}[b]{0.25\textwidth}
        \centering
        \includegraphics[width=\textwidth]{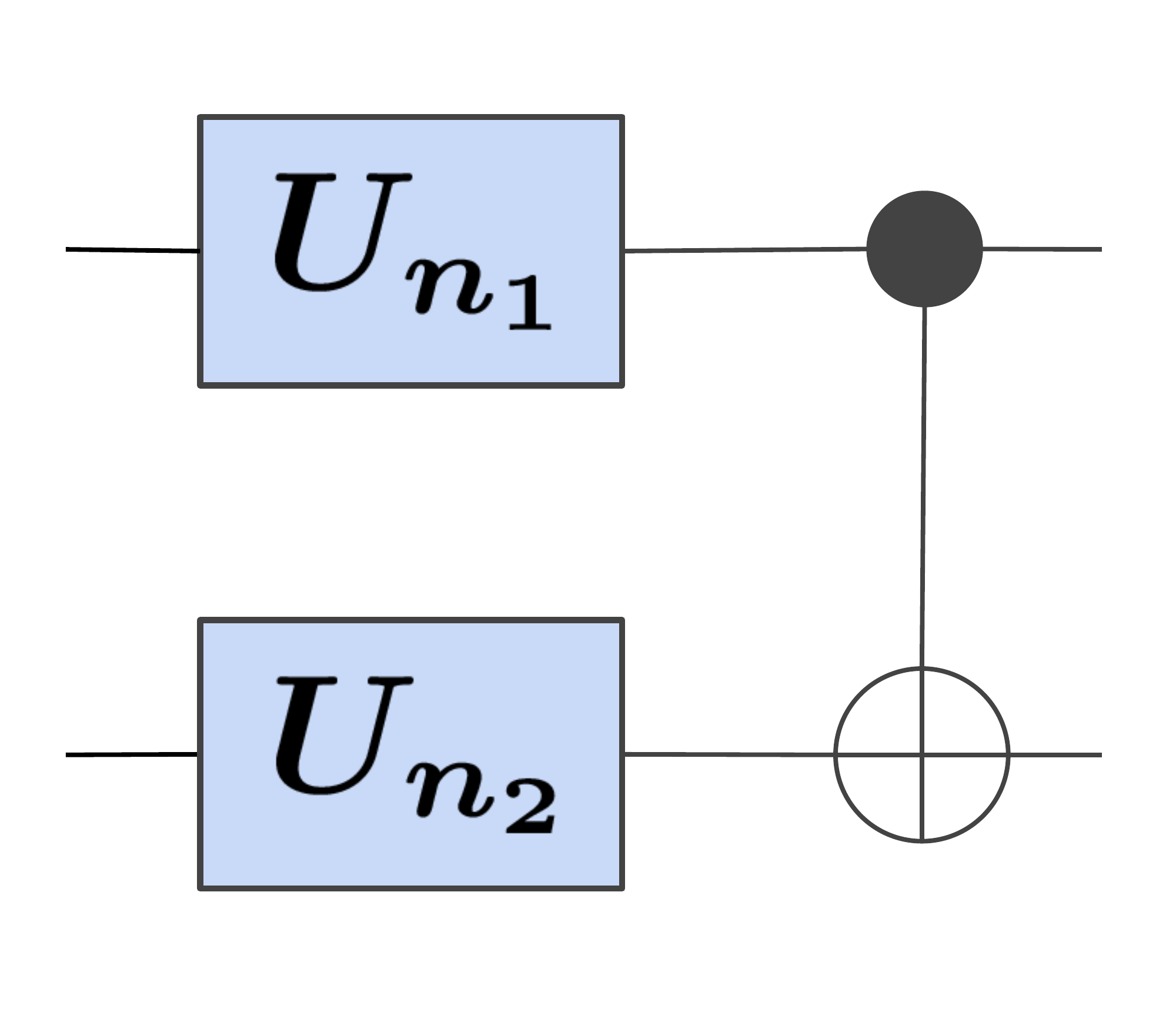}
        \caption{Simple gate}
        \label{fig:cnot}
    \end{subfigure}
    \hfill
    \begin{subfigure}[b]{0.25\textwidth}
        \centering
        \includegraphics[width=\textwidth]{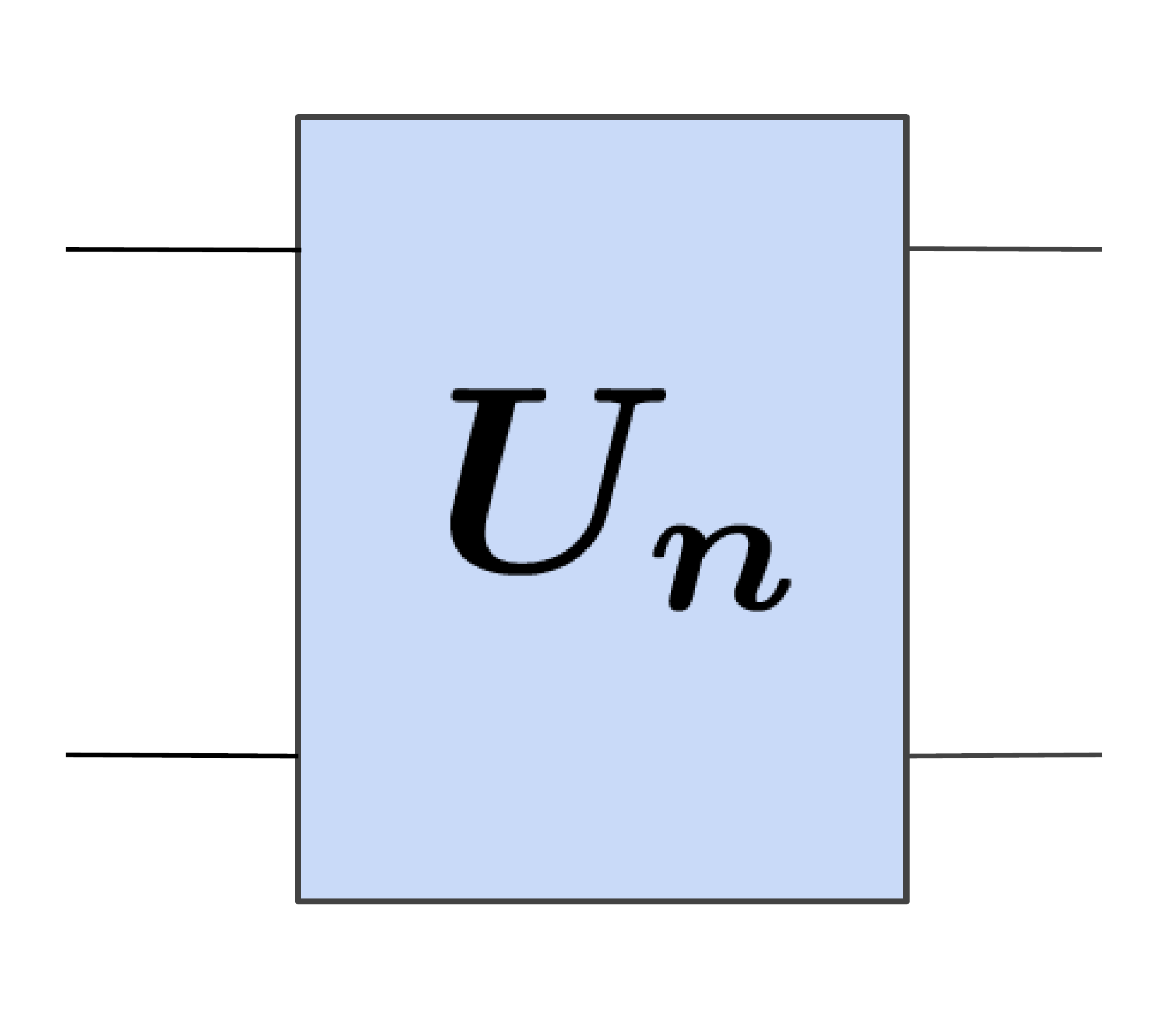}
        \caption{General gate}
        \label{fig:arbitrary}
    \end{subfigure}
    \hfill
    \begin{subfigure}[b]{0.25\textwidth}
        \centering
        \includegraphics[width=\textwidth]{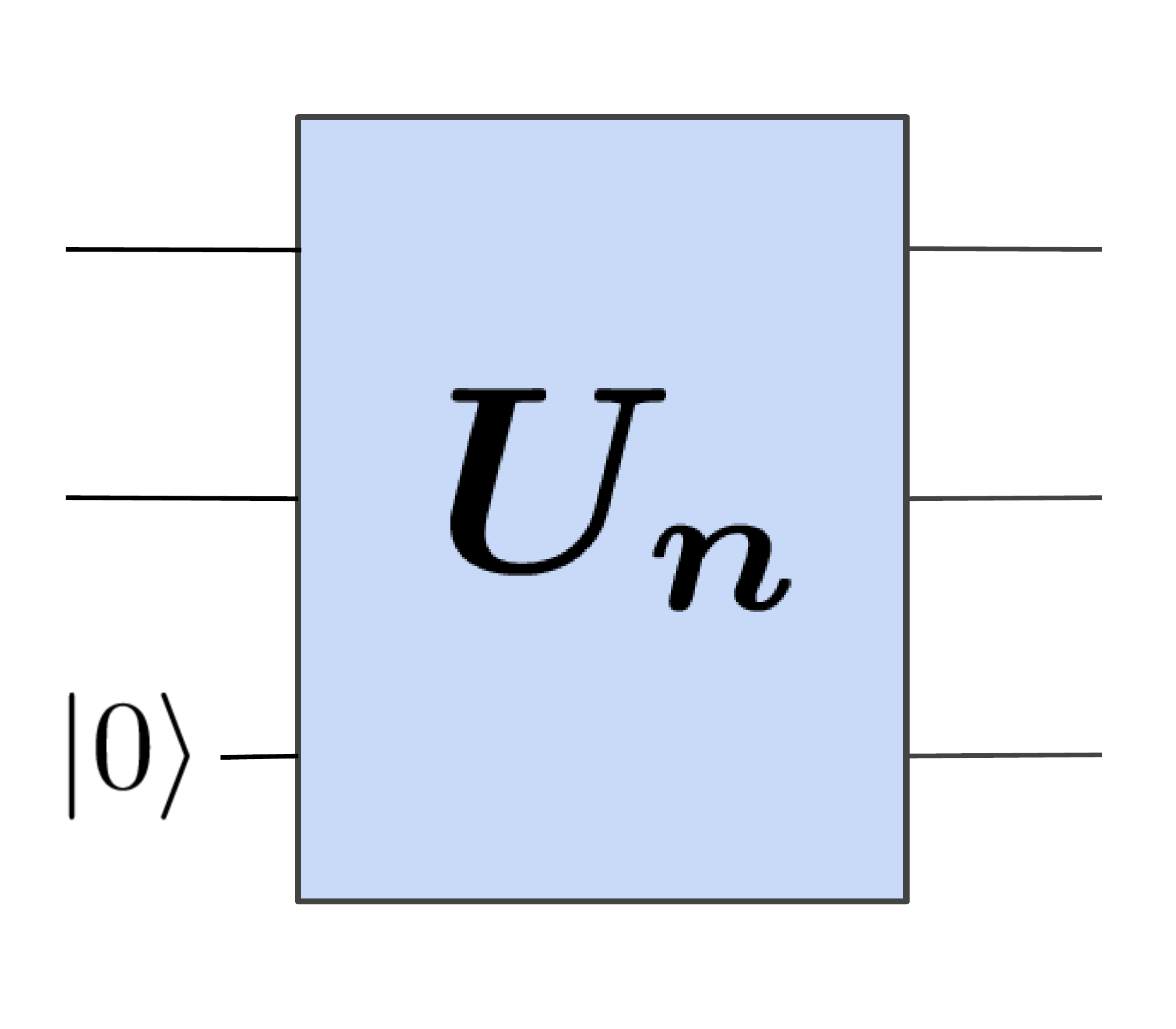}
        \caption{Ancilla gate}
        \label{fig:ancilla}
    \end{subfigure}
    \caption{\textit{Three alternative parameterizations of the unitary blocks in Fig~\ref{fig:networks}.}  (a) Two arbitrary single-qubit rotations followed by a CNOT. The direction of the CNOT may be reversed to preserve the causal structure of the network. This simple setting can be readily implemented in available quantum computers. (b) An arbitrary two-qubit gate. Such general setting would in practice require compilation into low-level hardware-dependent gates. (c) An arbitrary three-qubit gate involving an ancilla qubit. The ancilla is traced out allowing to perform a rich set of non-linear operations. Implementation of the latter in currently available hardware would require a compilation step.}
    \label{fig:three_graphs}
\end{figure}

We also explored a number of other methods for parameterizing the unitaries; Fig~\ref{fig:three_graphs} illustrates three such paramaterizations. In Fig.~\ref{fig:three_graphs}~(a), the unitary block is composed of two arbitrary single-qubit rotations and a CNOT$_{ij}$ gate, where $i$ and $j$ are control and target qubit, respectively. Note that in some cases the direction of the CNOT$_{ij}$ may be reversed in order to respect the causal structure. For example, in our 8 qubit implementations we reverse the control and target qubits for blocks $U_{2}$, $U_{4}$ and $U_{6}$ lying in the lower part of the circuit. In the case of the restriction to $SO(4)$ the single-qubit rotations are simply $Y$-rotations. 

In Fig.~\ref{fig:three_graphs}~(b), the unitary block consists of an arbitrary two-qubit gate. It is interesting to explore this much more general setting in simulations, although a practical implementation of such unitary may be costly. That is, the two-qubit unitary needs to be compiled into low-level hardware-dependent gates. 

Finally, Fig.~\ref{fig:three_graphs}~(c) shows a three-qubit gate involving an ancilla qubit. By tracing out the ancilla qubit we can effectively implement a rich class of non-linear functions, e.g. step functions~\cite{wan2017quantum}, closely resembling the operations of classical neural networks. Again, in practice a significant overhead is expected due to compilation. 

The measurement $\hat{M}$ is performed on a specific qubit and consists of a simple Pauli measurement in a chosen direction. This can be implemented in practice by an additional single-qubit rotation followed by the projective measurement onto $|0\rangle\langle 0|$. This is sufficient for a binary classification task; by computing and thresholding the expectation value of $M$, TTN and MERA classify the input $\psi^d$ into one of the two classes. In our example in Fig.~\ref{fig:networks}, the measurement is performed on qubit number six.

\subsection{Learning process and complexity}

We now discuss the learning process. In principle, the circuit parameters would be adjusted to directly maximize the classification accuracy on the training set or, in other words, minimize the classification error. Optimizing such an objective function is highly non-trivial and it is common to optimize a bound instead. Here we choose to minimize the mean square error between predictions and true class labels
\begin{equation}
J(\boldsymbol{\theta}) = \frac{1}{D}\sum_{d=1}^D \Big ( M_{\boldsymbol{\theta}}(\psi^d) - y^{d} \Big )^2 ,
\label{e:cost}
\end{equation}
where $\psi^{d}$ are inputs, $y^{d}$ are class labels, $D$ is the number of training data points, and $\boldsymbol{\theta}$ groups all the adjustable parameters of the circuit as described above. Although there exist several approaches to carry out this optimization, artificial neural networks are commonly optimized by stochastic gradient descent algorithms. At each iteration $t$, we estimate the gradient $\nabla J^{(t)}$ and choose a learning rate $\eta^{(t)}$. Parameters are then updated via a rule of the kind $\boldsymbol{\theta}^{(t+1)} \leftarrow \boldsymbol{\theta}^{(t)} + \eta^{(t)} \nabla J^{(t)}$. This algorithm is stochastic because at each iteration the gradient is  estimated on a small batch rather than on the full training set. Beside speeding up the calculation, this noisy gradient may help in escaping from local minima. Much literature and experimentation has been dedicated to improving stochastic gradient descent algorithms. In this work, we employ a variant called Adaptive Moment Estimation (Adam)~\cite{kingma2014adam}.

The cost function is a function of the measurement outcome of the circuit being trained. In order to obtain these measurement outcomes, the circuit itself must be evaluated. In Table~\ref{tab:complexity} we summarize the complexity of obtaining the measurement outcomes at the end of the different types of circuits in this paper. The complexity stated is in terms of the number of multiplications of scalar numbers required to perform the task. The complexities in the two dimensional cases are stated for a grid of $N \times N$ qudits. The complexities stated for the two-dimensional networks use the network architecture introduced in Refs.~\cite{2007PhRvL..99v0405V} and~\cite{vidal2008class}. 

In the case of efficiently contractable networks we can compute the exact gradient using off-the-shelf automatic differentiation software (e.g., TensorFlow~\cite{abadi2016tensorflow}). This applies to many one-dimensional networks including TTNs and MERA. For networks that cannot be efficiently contracted a finite-difference method or an approximation to the true gradient must be used~\cite{farhi2018classification}. These strategies introduce additional noise due to finite-sampling error, and intrinsic noise of near-term quantum devices. We begin exploring the impact of these with simulations in Sec.~\ref{s:noise}. Note that all of the circuits we train in this paper can be evaluated efficiently on quantum hardware.

\begin{table}[!htb]
\centering
\begin{tabular}{llll}
\textbf{Dimension} & \textbf{Classifier} &\textbf{Qubit encoding} & \textbf{Amplitude encoding}\\ \hline
1D & TTN &  $O(N\chi^5)$ &  $O(N\chi^{N+5})$ \\
& MERA &  $O(N^{6\log_2{\chi}} \chi^4)$ &  $O(N^{6\log_2{\chi}} \chi^{N+4})$ \\
\midrule     
2D & TTN & $O(N^2\chi^5)$  & $O(N^2\chi^{2N+5})$  \\
& MERA & $O(\chi^{11N-16})$  & $O(\chi^{13N-16})$  \\
\end{tabular}
\caption{\textit{Computational complexity of hierarchical quantum classifiers under different data encoding.} The complexities indicate the number of multiplications of scalar numbers required to obtain the measurement outcomes. We use $N \chi$-dimensional qudits in one dimension and $N \times N$ $\chi$-dimensional qudits in two dimensions.}
\label{tab:complexity}
\end{table}

\subsection{Experimental results: Iris dataset}\label{sec:experiments}

In this experiment we tested the ability of a TTN to classify varieties of Iris. The Iris dataset~\cite{fisher1936use} consists of $150$ examples in total of three varieties of Iris flowers. Each example of Iris is described by four real valued attributes $x_{1-4}$. We encoded the four attributes into four qubits using Eq.~\eqref{e:encoding}. We then parameterized unitaries using the simple gate shown in Fig~\ref{fig:three_graphs}~(a), and restricted the single-qubit rotations to be real (i.e., Y-rotations). To allow for binary classification, three binary datasets were extracted from the original set. In each subset, each class comprised $1/2$ of the examples. For each class, $1/3$ of examples were used as a test set and used to compute the accuracy. Mean accuracy and one standard deviation computed on five random initializations are given by Table~\ref{iris-pid}. As shown, TTN performed extremely well in all cases.

\begin{table}[!htb]
\centering
\begin{tabular}{llllll}
\textbf{Classifier} & \textbf{Unitaries} & \textbf{Rotations} & \textbf{1 or 2} & \textbf{2 or 3} & \textbf{1 or 3} \\ 
\hline
TTN & Simple & Real & $100.00\pm 0.00$ & $96.77\pm 0.00$ & $100.00\pm 0.00$  \\                                          
\end{tabular}
\caption{\textit{Binary classification accuracy on the Iris dataset.} Mean test accuracy and one standard deviation are reported for TTN classifiers with five different random parameter initializations. The Iris dataset consists of three classes. From these we constructed three binary classification tasks.}
\label{iris-pid}
\end{table}

\subsubsection{Experimental results: Handwritten digits (MNIST)} 

In this experiment we tested the ability of TTN and MERA classifiers on a number of handwritten digit recognition tasks and compared the performance of different parameterizations. MNIST~\cite{lecun1998mnist} is a canonical data-set consisting of $70,000$ labelled gray-scale images of handwritten digit from $0$ to $9$. From this dataset we generated four binary classification tasks. In the first we kept only images containing $0$ or $1$, and for the second task, only $2$ or $7$. For the third tasks we re-labelled all images as even or odd. For the final task we divided the images into those that were greater than $4$ or not. MNIST images are $28\times28$ pixels. To allow for simulation using $8$ qubits, we performed principal component analysis on the images for each task and kept only the $8$ components with highest variance. Finally, we used Eq.~\eqref{e:encoding} to encode the data.

\begin{figure}[!htb]
    \centering
    \includegraphics[width=.55\textwidth]{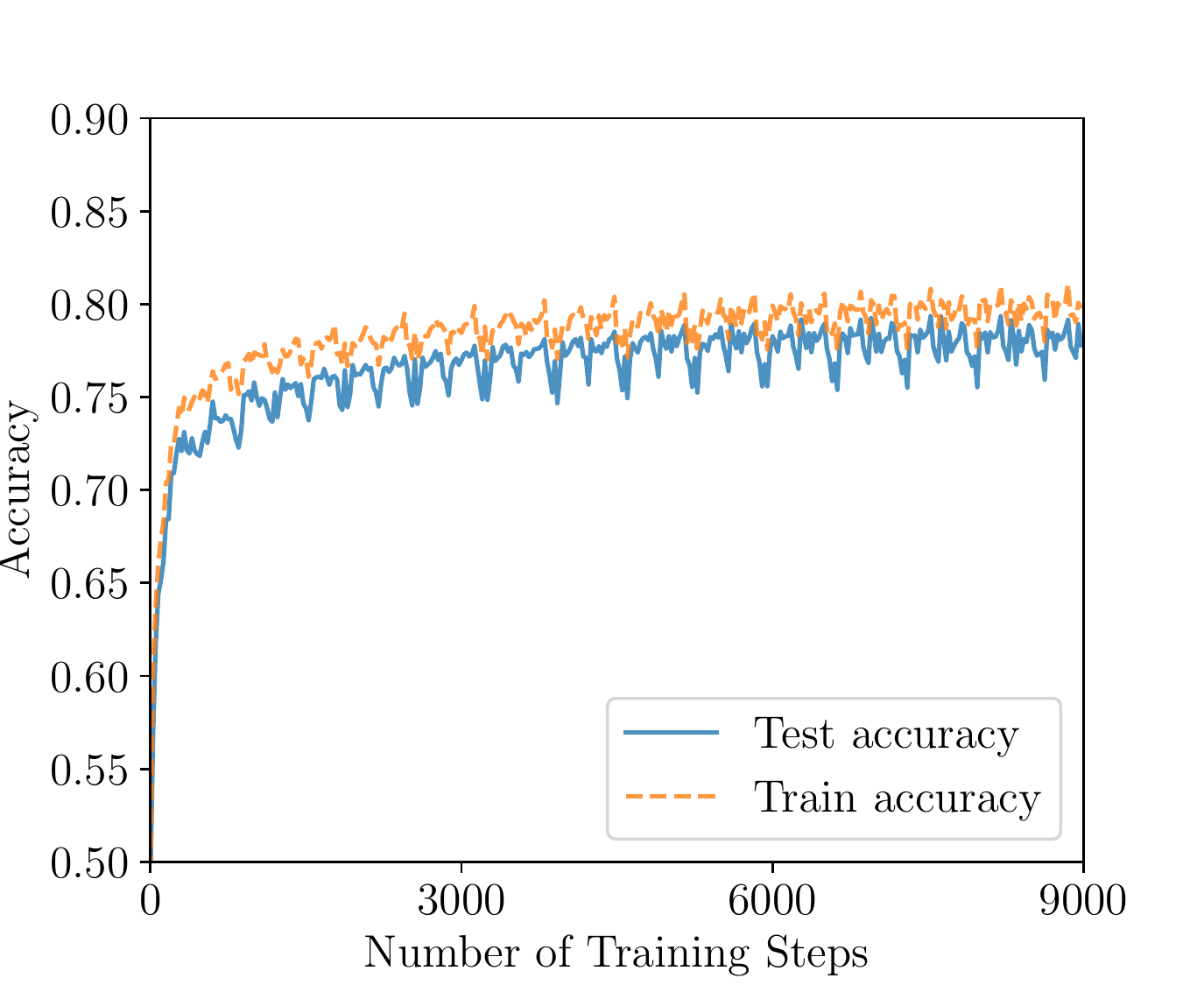}
    \caption{\textit{Train and test accuracy versus number of training steps.} Here we show typical results for a MERA classifier parametrized using general gates and complex rotations, applied to the \mbox{``Is $>4$''} task on the MNIST dataset with the dimension of each example reduced to $8$ using PCA.}
    \label{fig:training-MNIST-acc}
\end{figure}

Of the $70,000$ examples $55,000$ were used for training, $5,000$ for validation and $10,000$ for testing. Training was performed using the Adam optimizer~\cite{kingma2014adam} with batches of $20$ examples. Validation and test accuracy were recorded every $10$ training batches, and training was stopped when validation set accuracy did not increase for $30$ consecutive tests. Figure~\ref{fig:training-MNIST-acc} shows typical learning curves for train and test datasets.

Mean accuracy and one standard deviation computed on five random initializations are given by Table~\ref{tab:MNIST-acc}. The `Classifier' column describes if the circuit was a TTN, MERA or hybrid, that is, a MERA pre-trained with TTN. The `Unitaries' column describes if the circuit was parameterized using a simple, general or ancilla gate set as described by Fig.~\ref{fig:three_graphs}. The `Rotations' column specifies the type of rotation used, either real, $SO(4)$, or complex, $SU(4)$.

Some remarks are in order. First, we note that the restriction to simple unitaries led to significantly lower accuracy than when using general unitaries. Complex rotations improved the accuracy of the classifiers in all tasks except for task `0 or 1' where accuracy was already $>99.5\%$ with real rotations. It is notable that this is the case despite the input data being real-valued. Second, the MERA classifiers achieved higher accuracy than TTN classifiers in all cases, demonstrating the power of the additional unitaries. Third, the hybrid classifier achieved accuracy comparable to that of the standard MERA. On average, hybrid classifiers required $2.45$ times more training steps until convergence than standard MERA. However, the number of post-training steps required was only $0.825$ times the number of training steps of standard MERA. This indicates that classical pre-training may lead to a reduction in the number of training steps carried out on the quantum computer, a potential advantage in the near-term. Finally, all networks outperformed the logistic regression benchmark except those using the simple gate-set.

One may wonder whether the accuracy on some of the tasks can be made more competitive with the state-of-the-art results on MNIST. In order to efficiently simulate all the circuits, each $28\times28$ image was reduced to an $8$-dimensional vector using PCA, thereby discarding a lot of information that could be useful for classification. To verify this we ran logistic regression without PCA on the most difficult of the four tasks, \mbox{``Is $>$ 4''}. This model achieved a test accuracy of $87.09\%$, a significant improvement over the logistic regression on the PCA reduced data which achieved $70.7\%$ instead (Tab.~\ref{tab:MNIST-acc}). We concluded that reducing the dimensionality of the data can have a detrimental effect on the model accuracy and therefore we expect TTN and MERA classifiers to perform better when using more principal components, or even raw data.

\begin{table}[!htb]
\centering
\begin{tabular}{lllllll}
\textbf{Classifier} & \textbf{Unitaries} & \textbf{Rotations} & \textbf{Is} $\mathbf{>4}$&\textbf{Is even}& \textbf{0 or 1} & \textbf{2 or 7} \\ \hline
TTN & Simple & Real &$65.59\pm0.57$&$72.17\pm0.89$&$92.12\pm2.17$&$68.07\pm2.42$\\
TTN & General & Real &$74.89\pm0.95$&$83.13\pm1.08$&$99.79\pm0.02$&$97.64\pm1.60$\\
MERA & General & Real &$75.20\pm1.51$&$82.83\pm1.19$&$99.84\pm0.06$&$98.02\pm1.40$\\
Hybrid & General & Real &$76.30\pm1.04$&$83.53\pm0.21$&$\mathbf{99.87}\pm0.02$&$98.07\pm1.46$\\
TTN & Simple & Complex  &$70.90\pm0.73$&$80.12\pm0.64$&$99.37\pm0.12$&$94.09\pm3.37$\\
TTN & General & Complex  &$77.56\pm0.45$&$83.53\pm0.69$&$99.77\pm0.02$&$97.63\pm1.48$\\
MERA & General & Complex    &$\mathbf{79.10}\pm0.90$&$\mathbf{84.85}\pm0.20$&$99.74\pm0.02$&$\mathbf{98.86}\pm0.07$\\
Hybrid & General & Complex &$78.36\pm0.45$&$84.38\pm0.28$&$99.78\pm0.02$&$98.46\pm0.19$\\
Logistic & N/A & N/A &$70.70\pm0.01$&$81.72\pm0.01$&$99.53\pm0.01$&$96.17\pm0.01$\\
\end{tabular}
\caption{\textit{Binary classification accuracy on the MNIST dataset.} Mean test accuracy and one standard deviation are reported for TTN, MERA and hybrid classifiers with five different random initial parameter settings using two different types of unitary parametrization. Hybrid classifiers consist of pre-training a TTN classifier and that transforming it into a MERA classifier by training additional unitaries.}
\label{tab:MNIST-acc}
\end{table}

\subsection{Experimental results: Quantum data}

We now consider the problem of classifying quantum data, that is, quantum states generated by different physical processes. A physical process can be simulated by a quantum circuit. By setting up two different quantum circuit layouts, we can generate synthetic classification tasks. Let us first define the \textit{building block} for our quantum circuit layouts. 

Our building block consists of single-qubit rotations $U_i$ for all qubits $i\in\{0,\dots,N\}$, followed by all the possible CNOT$_{ij}$ gates where $i$ and $j$ are control and target qubits, respectively, and $i<j$. The angles of the single-qubit rotations are the only parameters of our building block. By stacking several of these building blocks, we can generate deeper and more complex circuits. In particular, we chose to identify the class with the number of building blocks in the stack (e.g., class 5 consists of 5 building blocks). 

Now, for each class, we can generate a quantum state by randomizing all the single-qubit gates, and then executing the circuit on initial state $\ket{0}$. This is repeated many times in order to generate a dataset. As discussed in Sec.~\ref{sec:main_section}, we assume that each quantum state in the dataset can be directly fed into the quantum computer where the classifier is executed, hence not requiring any pre-processing. The tasks of the classifier is to determine which of two circuit layouts a state was generated from. 

\begin{figure}[!htb]
    \includegraphics[width=\textwidth]{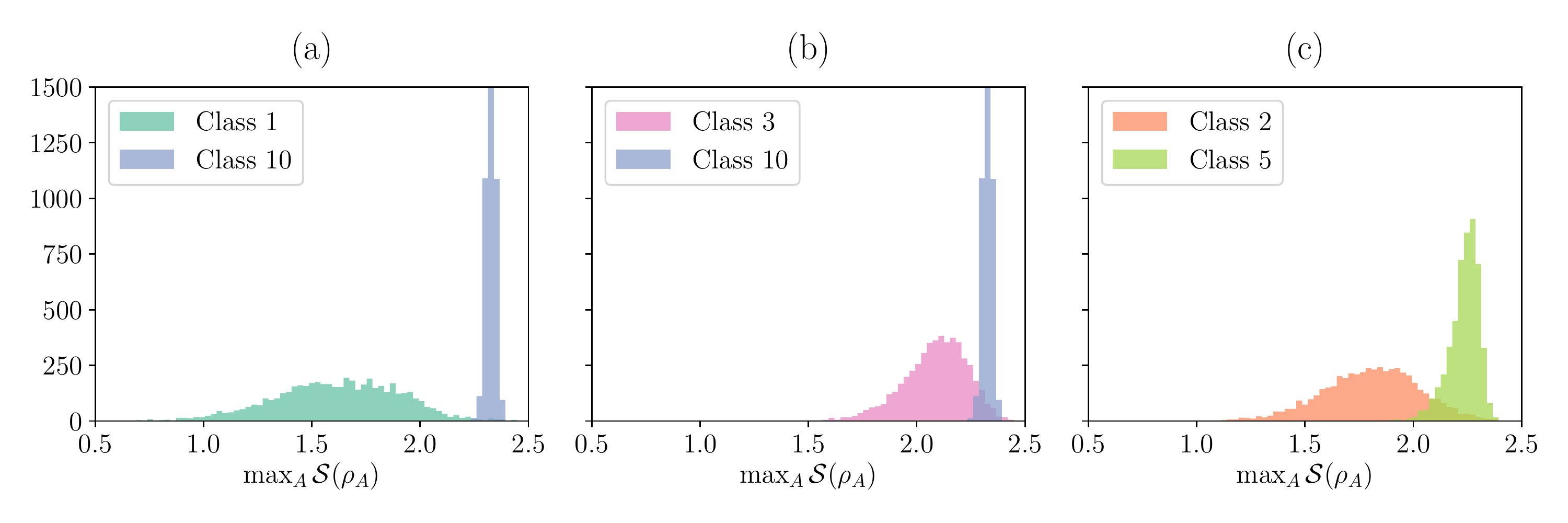}
    \caption{\textit{Distribution of the maximum bipartite entanglement entropy for synthetic quantum datasets.} Quantum data points were generated by random circuits with different number of building blocks $y \in \{1,2,3,5,10\}$ as explained in the main text. From this data we created three classification tasks: (a) 1 vs 10, (b) 3 vs 10, and (c) 2 vs 5. The subplots show histograms of maximum bipartite entanglement entropy for the three classification tasks. Such property could be used to separate classes and classify data with high accuracy, hence the synthetic classification tasks are well-posed. We stress that the computation of such property is intractable and do not expect the hierarchical classifiers to be able to exploit it when classifying input data.} 
    \label{fig:max_ent_entropy}
\end{figure}

Here, we work with circuits of $N=8$ qubits. We generated datasets of $D=5,000$ quantum states for each of the classes $y \in \{1,2,3,5,10\}$. To make sure that the synthetic classification task was well defined, we first looked for a strategy that could correctly classify the states most of the time. For each state, we computed the maximum bipartite entanglement entropy, $\max_A S(\rho_A)=\max_B S(\rho_B)$, over all possible partitions $A, B$ of the 8 qubits. Figure~\ref{fig:max_ent_entropy} shows histograms of this quantity for three classification tasks. By inspecting the overlap of distributions we can find an optimal threshold that would classify states correctly most of the time. This shows that the classification task is meaningful. We would like to stress that this is an intractable strategy. The only purpose is to demonstrate that, in principle, there is a feature of the state that correlates with the class. The hope is that a hierarchical quantum classifier can find equally successful strategies in a tractable way.

The classifier used for this task was a TTN like the one shown in Fig.~\ref{fig:networks}. We considered two parameterizations; the first uses general gates such as the one shown in Fig.~\ref{fig:three_graphs}~(b). The second uses arbitrary three-qubit gates where one of the qubits is an ancilla initialized in the state $\ket{0}$, as illustrated in Fig.~\ref{fig:three_graphs}~(c). The data described above was divided into training, validation and test sets. Each of these sets were balanced, that is, they had an equal number of states from each class. A set of $1000$ examples from each class was held out as a test set. Training was performed for $4,000$ iterations with batches of $40$ states and test accuracy was recorded every $50$ iterations. The best test accuracy was recorded for each task.

Table~\ref{tab:depth-acc} reports mean classification accuracy and one standard deviation computed on five random initializations. Results for the TTN with general two-qubit gates are no better than random class assignment in all tasks, indicating the need for a more expressive model. Indeed, when using gates augmented by an ancilla qubit, TTN was able to classify quantum states with some accuracy, suggesting that those may play a key role. The classification accuracy is higher for the `1 or 10' task; this is somewhat expected as the overlap of classes 1 and 10 shown in Fig.~\ref{fig:max_ent_entropy}~(a) is less than that of the other tasks shown in Figs.~\ref{fig:max_ent_entropy}~(b) and~(c).

Finally, as a proof of principle, we verified the performance of a classical logistic regression model. We fed the vector of amplitudes to the model and trained with off-the-shelf software. The test accuracy was close to $50$\%, that is, no better than random. We shall stress that this approach is not feasible in practice, since only providing the input in classical form would require full tomography of the quantum dataset.

\begin{table}[!htb]
\centering
\begin{tabular}{llllll}
\textbf{Classifier} & \textbf{Unitaries} & \textbf{Rotations} & \textbf{1 or 10} & \textbf{3 or 10}  & \textbf{2 or 5} \\ \hline
TTN & General & Complex & $50.25	\pm	0.58	$&$	49.95\pm	0.72	$&$	50.36	\pm	1.05	$ \\
TTN & Ancilla & Complex & $\mathbf{64.05}	\pm	1.12	$&$	\mathbf{59.33}	\pm	0.22	$&$	\mathbf{58.02}	\pm	0.65$\\
\end{tabular}

\caption{\textit{Binary classification test accuracy on synthetic quantum datasets.} Mean test accuracy and one standard deviation are reported for TTN classifiers with five different random initial parameter settings using two different types of unitary parametrization.}
\label{tab:depth-acc}
\end{table}

\subsection{Experimental results: Characterizing the effect of noise on classification performance}\label{s:noise}

Many machine learning models including neural networks are highly robust against the negative effects of noise. Some kinds of noise can even help with convergence and generalization \cite{graves2011practical,achille2017emergence}. In this experiment we tested the effect of depolarizing noise on the quantum classifier by simulating a depolarizing channel. It consists of a completely positive map $\Delta _{\lambda }$ parametrized by $\lambda$ from a \mbox{$2^{N}$-dimensional} state $\rho$ to a linear combination of $\rho$ and a maximally mixed state
\begin{equation}\label{e:depolarize}
\Delta _{\lambda}(\rho)=\lambda \rho +{\frac  {1-\lambda }{2^{N}}}I .
\end{equation}

We used one of the TTN classifiers for classes 1 and 2 of the Iris dataset (see Sec.~\ref{sec:experiments}) and simulated the noisy circuit using the IBM Quantum Experience. The depolarizing channel was applied to the system after the application of each unitary gate in the circuit, that is, after each single-qubit rotation and CNOT gate. The entire test set was used to evaluate accuracy.

In order to make a realistic case, we used a finite number of measurements to estimate the class predictions. For each data point, we took 401 measurements in the computational basis and obtained the most likely class by majority vote. The 401 measurements may not be sufficient to estimate the output of the circuit with high confidence when the probability assigned to both classes is close to 0.5. In other words, repeating the 401 measurements and taking the majority vote could lead to a different class assignment for the very same data point. Therefore, we repeated the computation of the accuracy 200 times and obtained error bars. Finally, we increased the amount of noise $\lambda$ from $0$ to $0.2$ in increments of $0.01$.

\begin{figure}[htb!]
    \centering
    \includegraphics[width=.55\linewidth]{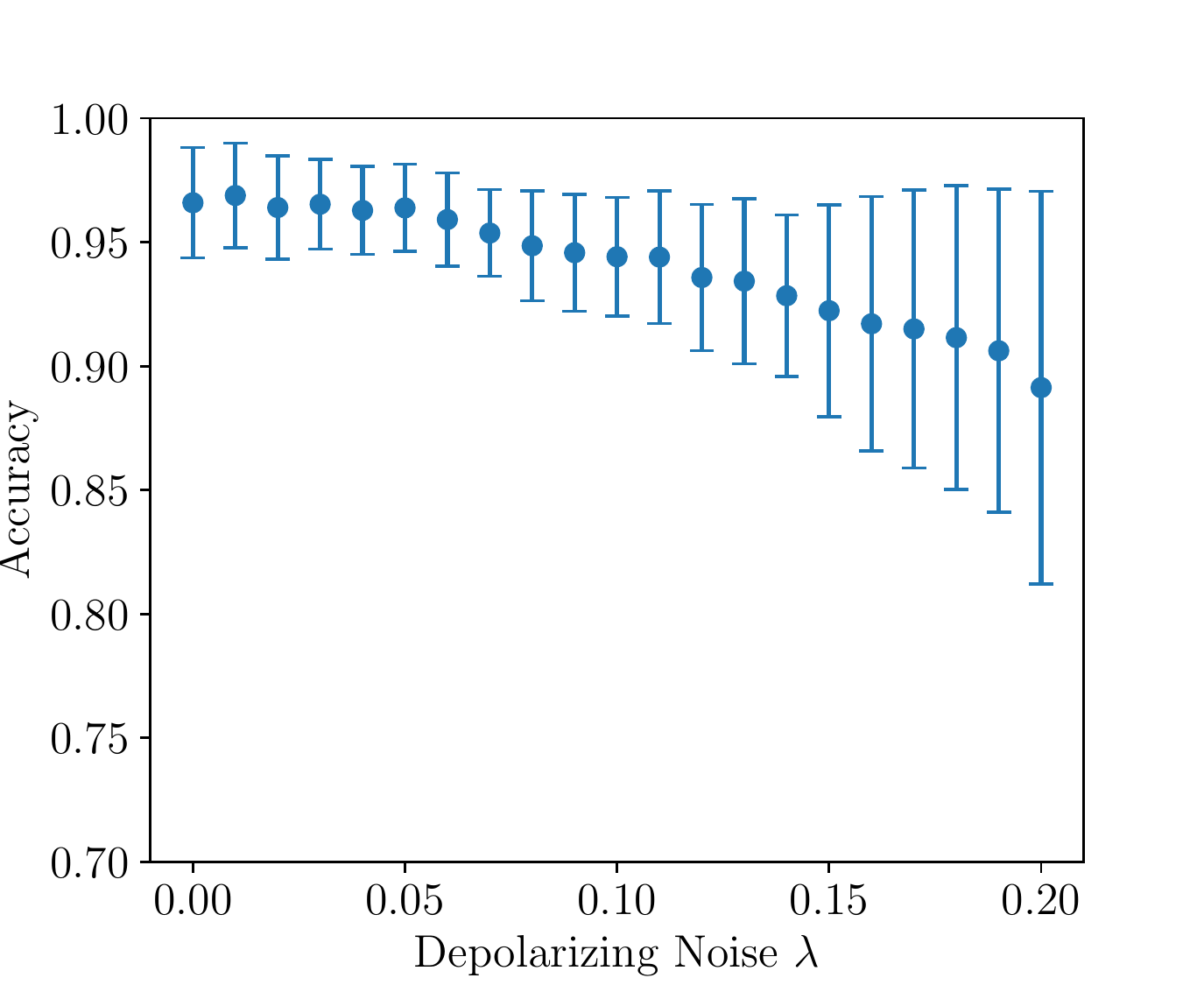}
    \caption{\textit{Effect of depolarizing noise and finite sampling noise on the accuracy of the TTN Iris classifier.} We show mean and one standard deviation of the classification accuracy computed on the test set. The mean accuracy remains above $95\%$ for depolarizing noise up to $\lambda = 0.07$ showing some level of resilience in the model. As we increase the depolarizing noise further, (i) the model gets worse and mean accuracy reduces, and (ii) the standard deviation increases indicating the need for more measurements to overcome the finite sampling noise.}
    \label{fig:noise}
\end{figure}

Figure~\ref{fig:noise} shows mean and one standard deviation of the classification accuracy on the test set. We first noticed that finite sampling led to some error even when no depolarizing noise was used. Indeed, we obtained a mean accuracy of 96.5\% with $\lambda=0$; the very same model achieved 100\% accuracy under exact computation (see results for ``1 or 2'' in Tab.~\ref{iris-pid}). Second, the mean accuracy reduced as we injected depolarizing noise, but it remained above $95\%$ for depolarizing noise up to $\lambda = 0.07$ showing some level of resilience. Finally, as we increased the noise further, the standard deviation of the accuracy increased as well. This is expected: as the output state gets closer to the maximally mixed state according to Eq.~\ref{e:depolarize}, the probability assigned to both classes gets closer to 0.5. Hence, a larger number of measurements would be needed to estimate the class.


\subsection{Experimental results: Deployment on a quantum computer}
\begin{figure}[!htb]
    \centering
    \begin{minipage}{0.5\textwidth}
    \includegraphics[width=\linewidth]{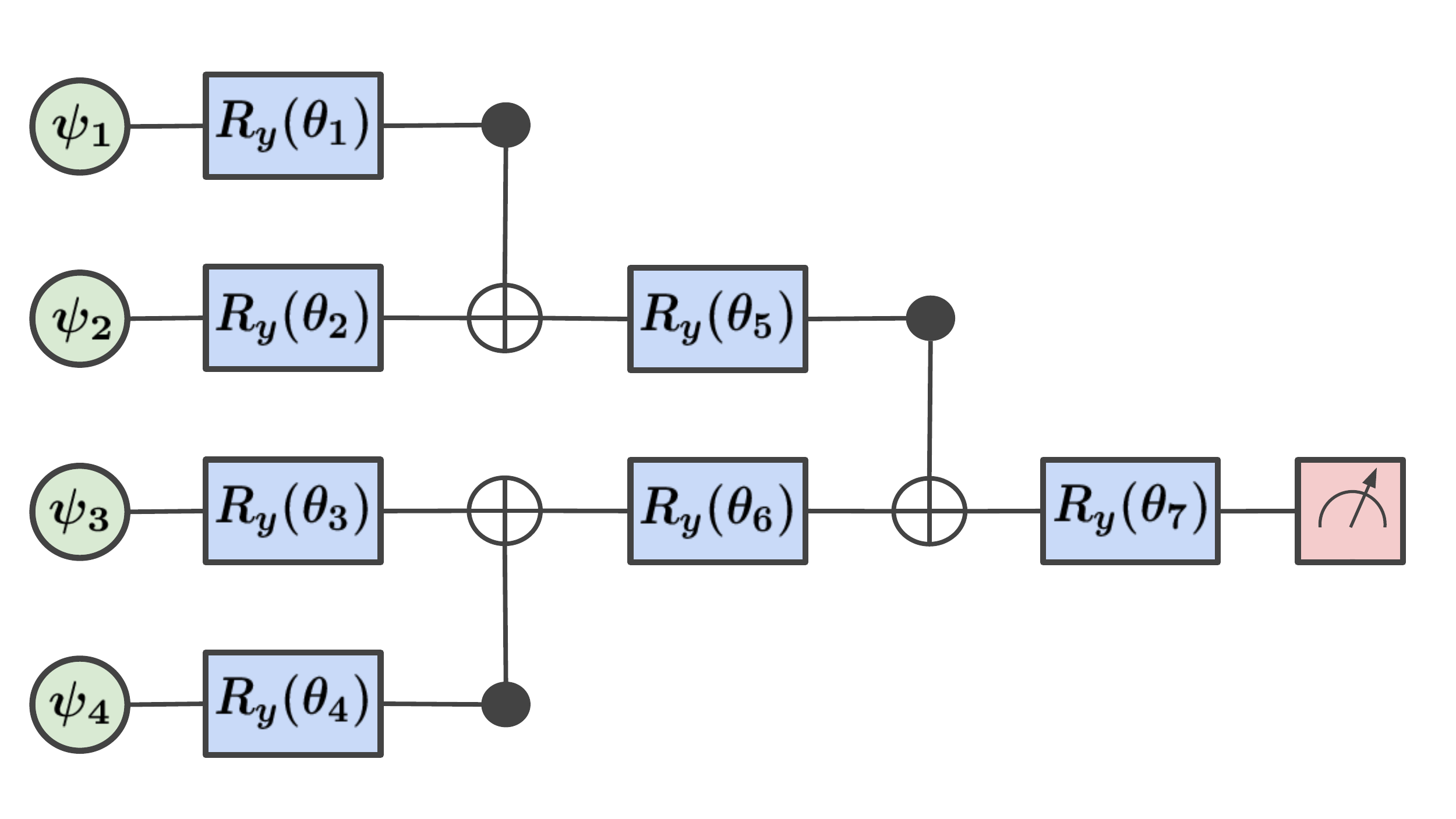}
    \end{minipage}
    \caption{\textit{Iris TTN classifier circuit schematic.} The TTN classifier uses a simple unitary parametrization with real rotations. It was trained classically and then deployed on the ibmqx4 quantum computer.}
    \label{fig:iris_circuit}
\end{figure}

In this experiment we deployed the Iris classifier for classes 1 and 2 (see Sec.~\ref{iris-pid}) on the ibmqx4 quantum computer available in the IBM Quantum Experience. As shown in Fig.~\ref{fig:iris_circuit}, this TTN classifier has three CNOT gates and seven rotations in the Y direction. A test set of $34$ unseen examples was used to determine accuracy. For each example, the circuit was run $401$ times, and the samples were used to compute the most likely class. The circuit correctly classified $100\%$ of the test set, and achieved a test cost function value of $0.0811$ (Eq.~\eqref{e:cost}).

\section{Discussion}\label{sec:discussion}

Combining the success of deep neural networks and other machine learning methods with the power of quantum computation is a tantalizing prospect. Much work to date has focused on modifying classical machine learning algorithms to incorporate quantum linear algebra subroutines, thus inheriting their speed-ups. One such subroutine is the quantum algorithm for solving linear systems, also known as HHL~\cite{lloyd2010quantum}. The algorithm is exponentially faster than the best known classical alternative, although this comes with some caveats~\cite{aaronson2015read}. Quantum classifiers that use HHL include the quantum support vector machine~\cite{rebentrost2014quantum} and the kernel least squares~\cite{schuld2016prediction}. Whilst promising, these algorithms also inherit the limitations of HHL, in particular, the requirement that classical data be efficiently prepared in amplitude encoding. Another quantum subroutine that can be readily embedded in a quantum classification model is Grover's algorithm which, for example, has been used to improve both computational and statistical complexity of the perceptron model~\cite{kapoor2016quantum}.

While the above proposals assume availability of universal quantum computers, much of the recent literature has been focusing on algorithms for noisy intermediate-scale quantum technologies~\cite{preskill2018quantum}. These consist of hybrid quantum-classical algorithms where the quantum computer is used to execute ansatz circuits and to measure observables of interest. A classical optimization routine is used to adjust the ansatz circuit in order to minimize a cost function. Originally proposed for quantum chemistry and combinatorial optimization, these approaches have been recently investigated for supervised~\cite{farhi2018classification} and unsupervised~\cite{otterbach2017unsupervised,benedetti2018generative} machine learning. The underlying ansatz circuits are often inspired by the structure of classical neural networks, but without explicit reference to tensor networks. 

References ~\cite{liu2017machine} and~\cite{2018arXiv180311537H} propose training tree-like tensor networks to be classifiers. In particular, Ref.~\cite{liu2017machine} demonstrates that TTNs can be used to classify images of handwritten digits and to encode classes of images in quantum many-body states. The framework proposed in Ref.~\cite{2018arXiv180311537H} examines the role of training TTNs to be classifiers in a quantum computing context and provides numerical evidence that TTNs can be used to perform supervised and unsupervised machine learning with the support of a quantum computer. Our work extends these ideas in a number of respects. Firstly, more complex networks such as MERA are studied and their superiority relative to simpler networks is demonstrated. Secondly, it is shown that these networks can be used to classify quantum mechanical data in addition to classical data. Thirdly, networks are demonstrated that are constructed from simple two-qubit gates that can be natively implemented on available hardware. Finally, a trained tensor network is successfully deployed on a real quantum device (ibmqx4).

\section{Conclusion}\label{sec:conclusions} 

In this report we have demonstrated that hierarchical quantum circuits can be used to classify classical and quantum data. Circuits based on the Multi-Scale Entanglement Renormalization Ansatz (MERA) outperform simpler tree-like circuits known as Tree Tensor Networks (TTN). These circuits can be parameterized with a simple gate set that can be easily implemented on existing quantum computers. A trained model is shown to be resistant to depolarizing noise and is successfully deployed on the ibmqx4 quantum computer. 

Both MERA and TTN are naturally extendable to larger inputs. In $1$D each additional layer doubles the dimensionality of the input. It is less clear how to increase or decrease the modelling power of a circuit without changing the dimension of the input. In classical neural networks this is achieved by increasing the depth and breadth of the network. One possibility for accomplishing this with quantum hierarchical classifiers is to use $\chi$-level quantum systems (qudits) for some suitable $\chi>2$ as the unit of computation, rather than qubits ($\chi=2$). Ref.~\cite{liu2017machine} demonstrates that model expressiveness in tensor network classifiers can be increased by increasing the input and internal bond dimensions. This is equivalent to performing computation using qudits. Data can be encoded in qudits using a generalization of qubit encoding described in Ref.~\cite{stoudenmire2016supervised}. Whilst it is possible to simulate qudits with qubits, there are practical considerations that can make this challenging~\cite{wang2015quantum}. 

Currently it is unclear what network architecture is ideal for a classification task, a thorough examination of the role entanglement plays in classification circuits may help illuminate this. Consider the case of a TTN circuit applied to a product state input. In this circuit the measurement qubit interacts with each other qubit in the circuit at most once and therefore its entanglement with the rest of the circuit will increase as unitaries are applied. If the measurement qubit is highly entangled with the rest of the network it will struggle to minimize the cost function Eq.~\eqref{e:cost} but clearly it is necessary to introduce some entanglement in the network for correlations between input qubits to be shared. Such a trade-off may limit the effectiveness of TTN circuits, especially as they are scaled to larger inputs. 

Constraining machine learning models using regularization can help them to generalize better to unseen data. Indeed, parameters with large magnitude are a characteristic of overfitting. The unitary constraint of quantum circuits naturally prevents parameters from becoming large, and it is likely acting as a strong regularizer. Additional regularization methods from the machine learning literature will become important in future quantum machine learning work. For example, the addition of noise during training of classical neural networks can also have a regularizing effect~\cite{graves2011practical} and help the model to learn invariant representations~\cite{achille2017emergence}. In our study, we did not simulate circuit noise during the training phase, but we did show high resistance to depolarizing noise during the prediction step. 

Much of the success of convolutional neural networks comes from their ability to learn layers of translation invariant representations using a shared set of weights. Translation invariance can be enforced in TTN and MERA by restricting the unitaries within each layer to be the same. Similarly, scale invariance can be enforced by restricting the unitaries between different layers of the circuit to be the same. The role of weight sharing in hierarchical quantum classifiers is a question for future research.

In this report we have identified two cases where the cost of classical simulation is thought to be exponentially harder than that on a quantum computer. The first of these, which we do not test, is when the hierarchical quantum classifier cannot be classically simulated even when the input is a product state, $2$D MERA circuits being one such example. The second case is when the input data consists of entangled quantum states. Here, an entirely classical approach may require expensive tomography and become intractable as the system size grows. While there are many existing methods for classifying $2$D classical data, developing methods for classifying quantum data is a promising research direction.

\paragraph*{Contributions.}
E.G., A.H., J.L., and A.G. designed the classifiers. M.B., E.G., and J.L. designed the experiments. A.H., and J.L. performed the complexity analysis. S.C., E.G., M.B., and J.L. wrote the code. E.G., and S.C. performed the experiments on ibmqx4. M.B., E.G., A.H., and  S.C. analyzed the experimental results. Supervision and guidance was provided by S.S., A.G., and V.S. All authors contributed to the final version of the manuscript.

\paragraph*{Acknowledgements.} 
E.G. is supported by the UK Engineering and Physical Sciences Research Council (EPSRC) [EP/P510270/1] and by Rahko Limited. M.B. is supported by EPSRC and by Cambridge Quantum Computing Limited (CQCL). S.S. is supported by the Royal Society, EPSRC and the grant ARO-MURI W911NF-17-1-0304 (US DOD, UK MOD and UK EPSRC under the Multidisciplinary University Research Initiative). J.L., and A.H. are supported by EPSRC [EP/L015242/1]. We gratefully acknowledge the support of NVIDIA Corporation with the donation of the Titan Xp GPU used for this research.

\paragraph*{Data availability statements.} 
All data needed to evaluate the conclusions are available from the corresponding author upon reasonable request.

\paragraph*{Conflict of interest.} 
The authors have no potential financial or non-financial conflicts of interest.

\bibliographystyle{unsrt}


\end{document}